\documentclass[acmsmall,screen]{acmart}

\AtBeginDocument{%
  \providecommand\BibTeX{{%
    \normalfont B\kern-0.5em{\scshape i\kern-0.25em b}\kern-0.8em\TeX}}}

\setcopyright{acmlicensed}
\copyrightyear{2024}
\acmYear{2024}
\acmDOI{XXXXXXX.XXXXXXX}

\usepackage{arabtex}
\usepackage{utf8}
\setcode{utf8}
\usepackage{tabularx}
\usepackage{footnote}
\makesavenoteenv{tabular}
\makesavenoteenv{table}

\usepackage{subcaption}

\newcommand{\change}[2]{\textcolor{black}{#1}\iffalse{#2}\fi}

\newcommand{\finalChange}[2]{\textcolor{black}{#1}\iffalse{#2}\fi}

\begin{document}

\title[hadith]{``The Prophet said so!'': On Exploring Hadith Presence on Arabic Social Media}


\author{Mahmoud Fawzi}
\affiliation{%
  \institution{School of Informatics, The University of Edinburgh}
  \country{UK}}
\email{m.f.g.ibrahim@sms.ed.ac.uk}

\author{Björn Ross}
\affiliation{%
  \institution{School of Informatics, The University of Edinburgh}
  \country{UK}}
\email{b.ross@ed.ac.uk}

\author{Walid Magdy}
\affiliation{%
  \institution{School of Informatics, The University of Edinburgh}
  \country{UK}}
\email{wmagdy@inf.ed.ac.uk}

\begin{abstract}
Hadith, the recorded words and actions of the prophet Muhammad, is a key source of the instructions and foundations of Islam, alongside the Quran. Interpreting individual hadiths and verifying their authenticity can be difficult, even controversial, and the subject has attracted the attention of many scholars who have established an entire science of Hadith criticism.  Recent quantitative studies of hadiths focus on developing systems for automatic classification, authentication, and information retrieval that operate over existing hadith compilations. Qualitative studies on the other hand try to discuss different social and political issues from the perspective of hadiths, or they inspect how hadiths are used in specific contexts in official communications and press releases for argumentation and propaganda. However, there are no studies that attempt to understand the actual presence of hadiths among Muslims in their daily lives and interactions. In this study, we try to fill this gap by exploring the presence of hadiths on Twitter from January 2019 to January 2023. We highlight the challenges that quantitative methods should consider while processing texts that include hadiths and we provide a methodology for Islamic scholars to validate their hypotheses about hadiths on big data that better represent the position of the society and Hadith influence on it.
\end{abstract}



\maketitle

\section{Introduction}
Religion continues to leave a remarkable footprint in public life \cite{ongaro} as well as the private lives of individuals \cite{moon}. Public topics where religion is still deeply involved include immigration \cite{bloom, cremer}, nationalism and populism \cite{shankar, haynes}, warfare \cite{hassner}, law \cite{bentzen}, and international relations \cite{sandal}. At the individual level, religion is considered while addressing topics like autism \cite{keri}, positive psychology \cite{davis}, mental health \cite{zong}, and accountability \cite{anderson}. 

Islam is the second most \change{followed}{#change7} religion in the world with around 24.9\% of the world population being Muslims \cite{cia}. It possesses a very special position as a cultural background that intervenes with different aspects of life. 4 out of the 193 members of the United Nations still define themselves as Islamic republics in their official names, namely the Islamic Republic of Pakistan, the Islamic Republic of Mauritania, the Islamic Republic of Iran, and the Islamic Republic of Afghanistan \cite{un}. The second largest intergovernmental organization after the UN is the Organization of Islamic Cooperation (OIC) having 57 member states \cite{hossain}. Many studies acknowledge and investigate the important role of political Islam during the past decade in the Arab spring that caused radical changes to the history of the Middle East \cite{esposito, ferrero}. Islam is also present when it comes to banking and finance with hundreds of Islamic banks operating all over the world \cite{kouser} and there has been a lot of research into the details and the rules of Islamic finance \cite{mk}.

The two major sources of the Islamic religion are the Quran, which is the holy book of Muslims, and the Hadith \cite{khan}. A hadith is a narration originating from the sayings and conduct of Muhammad, the prophet of Islam \cite{azmi2019computational}. These narrations, which are also collectively referred to as Hadith, are considered by some scholars to be the largest resource covering the Classical Arabic literature \cite{azmi2019computational}. \citet{azmi2019computational} believe this is valuable for modern applications because unlike English where old English, Middle English, and Modern English are considerably different, Classical Arabic on the other hand is a subset of Modern Standard Arabic that is being used nowadays for official Arabic communication. \finalChange{The Quran is universally recognized by Muslims worldwide as a singular text.}{#final3} In contrast, Hadith has many resources that vary in terms of their credibility, their compilation dates, and the background of the scholars who compiled them especially the branch of Islam they follow \finalChange{\cite{saloot}}{#final4}. The same incident of a hadith can very likely be reported by more than one resource where each of them might report a different variant from the others \change{in the wording of the prophet and the story narration}{#change9}. Consequently, although the number of authentic hadiths is around 30,000 \change{in some estimates}{#change17}, the total number of variants is in the order of hundreds of thousands \cite{jabbar}. More importantly, unlike Quran, it is permissible in Islam to narrate hadiths by meaning or paraphrase them \cite{nur}. This permission makes Muslims less sensitive to spelling mistakes when narrating hadiths. While the aforementioned aspects pose many challenges to processing hadiths, they also situate it as a rich, valuable, and dynamic language resource with many research opportunities.
Despite the vital role of Hadith in interpreting the guidelines driving political Islam, Islamic Banking, and other Islamic ideologies, there is little amount of quantitative research addressing Hadith \change{through a social lens}{#change10}. Attempts to build datasets or perform classification by topic or authenticity \change{ on classical Islamic books}{#change11} don't address the flexible and complex nature of Hadith. In this study, we aim to explore how social media users express themselves through Hadith and to introduce it to the research community taking into account the challenging aspects of its nature. The overarching goal of our research is to investigate how Muslims use hadiths on social media. On attempting to achieve this goal, three research questions naturally arise:

\begin{itemize}
    \item \textbf{RQ1:} What are the practical challenges of processing hadiths and covering the actual presence of hadiths among individuals?
    \item \textbf{RQ2:} Which hadiths are most frequently shared by users of Arabic social media?
    \item \textbf{RQ3:} What is the distribution of hadiths on Arabic social media in terms of topics, authenticity, time, and seasonality?
\end{itemize}

To address these questions, we extract a sample of \change{around 300K}{} tweets that include hadiths from the Twitter archive collection\footnote{\url{https://archive.org/download/archiveteam-json-twitterstream}} \change{covering a random sample of tweets between January 2019 and January 2023}{}. We also consult experts in Islamic studies for a source of high-level topical categorization of hadiths. We parse the most recommended source, namely sonnaonline\footnote{\url{http://www.sonnaonline.com}}. We attempt to precisely match tweets including hadiths, or \change{supposed}{#change12} hadiths, to \change{recognized hadiths found in the literature}{#change12}. This results in a rich dataset on the use of hadiths in social media posts, which we analyze. 


\change{Our work contributes to the literature of CSCW and HCI research studying different aspects of spiritual and religious presence online \cite{wolf2024still}. While most of the work in this area focuses more on the Western communities, our study here adds to the limited work that studies the Muslim and Arab communities \cite{al2021atheists,albadi2019hateful,albadi2022deradicalizing,ibtasam2019my,abokhodair,rifat2022putting, rifat2022situating}, which should help in further understanding these communities in a step for building more diverse and inclusive technologies.
Particularly, }{#change20} our work paves the way for the CSCW community to explore multiple research directions that are related to the actual presence of hadiths among Arabic-speaking Muslims in everyday interactions on social media as opposed to classical Islamic books. 
\change{The findings of our analysis highlight several interesting social dynamics and online collaborative behaviours by Muslims in the Arab world, which should inspire building and designing technologies by the CSCW/HCI communities.}{#change20}
In addition, our proposed machine learning tasks and datasets that include hadiths \change{would bring attention of researchers}{#change0} to tackle problems such as hadith identification and part-of-hadith extraction. We also anticipate that big data analysis \change{would have a bigger role complementing the existing qualitative}{} work in social studies related to hadiths.    


\section{Background \& Related Work}
In this section we present earlier works exploring religious presence on social media, \change{especially within the CSCW and social computing context}{}. Afterwards, we introduce a brief background about Hadith and its significance by presenting some studies that examined some social phenomena through the lens of Hadith. Finally, we go through works attempting to process hadiths computationally.

\change{\subsection{HCI Studies on Religion and Spirituality}}{#change21}

\change{
A recent survey study by \citet{wolf2024still} that there has been a 
slight increase in the number of HCI papers addressing religion and spirituality during the past decade, especially in the era of social media. Nevertheless, they highlight that HCI still has many gaps in this area that requires attention from the HCI and CSCW communities \cite{wolf2024still}. They show that designing religious or spiritual experiences has attracted more attention than focused studies which contribute knowledge. This shift towards experimental studies leaves some relevant real-world topics behind causing a general lag in this research direction.}{}
In \change{some}{#change9} studies, religion is interlinked with \change{extraneous}{#change9} constituents of social media and the digital world like memes \cite{bellar2013reading}, emoticons \cite{stanton2013islamic}, and even communities of gamers \cite{steffen2014playing}. Such examples show that updated and continuous examination and research about \change{techno-spirituality}{#change6} is necessary for a better understanding of different societies and human behavior.
\change{
In the following, we summarize some of the work in HCI and CSCW that targeted the Muslim and Arab communities.
}{}

\citet{hashmi} review studies that explore the representation of Islam on social media and conclude that 75\% of them are qualitative. They also show the main themes of these studies and demonstrate that they represent negative images of Islam more than positive ones. \citet{abokhodair} study the presence of Quran verses on social media quantitatively and interview a sample of users to inspect their motives behind sharing them. They find that verses on topics such as jihad are shared much less often \change{by Muslims}{#change13} than those on topics linked to worship, contradicting some media depictions of Muslim social media use and practice.
\change{They demonstrated that online religious expression in the form of Quran sharing extends offline religious life and supports new forms of religious expression that should be adapted by platforms' designers.}{}

\change{There has been a substantial amount of work exploring religion and content moderation.}{} Hate speech \change{in particular}{} is an ongoing challenge on social media that on many occasions has had a religious nature \cite{froio, hanzelka, albadi, albadi2022deradicalizing, albadi2019hateful}. In their literature review about online hate speech, \citet{castano} analyzed 67 papers, 33 of which were related to religious hate speech. They concluded that Islam is the most attacked religion in the world.
\change{\citet{albadi2019hateful,albadi2022deradicalizing} studied the characteristics of religious hate speech on Twitter as well as YouTube. They found that only 11\% of such content on Twitter is generated by bots. They also found that videos with such content are prevalent in search results and first-level recommendations of Youtube. \citet{darwish2018predicting} and \citet{evolvi} studied the Islamophobic content on social media after Paris attacks in 2015 and after Brexit respectively.}{#change24}
Similarly, misinformation was also present with a religious flavor as \citet{muhammed} show that some fake news in India linked the outbreak of COVID-19 to some religious groups. Social media is also a stage for the conflict between the clergy of different religions and the individuals. This \change{is}{#change0} perceived by \citet{solahudin2019internet} as a discord between religious populism and religious authority after applying it to the case of Muslims in Indonesia. 

\change{
A recent work by \citet{ibrahim2024tracking} investigated the interplay between health and religious practices among Muslim women and its challenging and conflicting nature. This knowledge reveals a lot of design considerations for tracking technologies.
\citet{ibtasam2019my} studied the social factors affecting low-income Muslim women's use of technology and concluded that applications designed for such women should consider seeking wider acceptance from their families and social network as well as the religious norms.
Another work by \citet{al2021atheists} studied religious polarization in Arab online communities and identified four different groups of users in terms of their attitude towards religion. They hypothesize that the engagement of users from a specific group to a given discussion can be more constructive than that of users from another group. This can be a valuable input to recommender systems that sort news feeds for users.}{}

\change{
We believe that these earlier CSCW works were essential and complementary to other CSCW studies that focused on designing religion friendly platforms \cite{kim2022social, markum2023designing} as well as those inspired by religious practices to reframe sustainable design \cite{rifat2020religion}. We hope that covering an understudied aspect of religion on social media like Hadith will add to this body of literature.\\
}{#change21}

\subsection{Hadith and its Significance}

\begin{figure}
\centering
\includegraphics[width=0.6\linewidth]{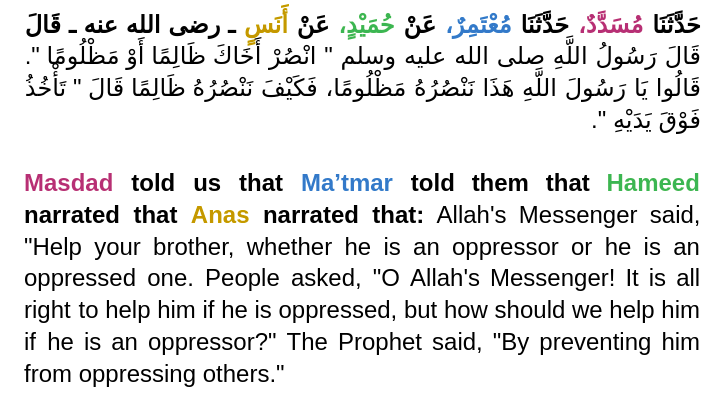}
\caption{A sample hadith obtained from \url{sunnah.com} with the English complete chain of narrators as reported by \url{qaalarasulallah.com}. \textbf{Isnad} in bold.}
\label{fig:hadith_sample}
\end{figure}

The precise definition of Hadith varies based on the branch of Islam as Sunni, Shia, and \change{Ibadi}{#change0} Muslims consider different collections of Hadith as their trusted resources for this vital scripture. In our study, we mainly focus on the Sunni standard since it is followed by the majority of Muslims, and in particular by most Muslims whose first language is Arabic. According to this standard, a hadith is any speech, discussion, action, approval, and physical or moral description attributed to the Prophet Muhammad, whether supposedly or truly \cite{saloot}. Scholars classify hadiths by their authenticity as \textit{authentic}, \textit{good}, \textit{weak}, or \textit{fabricated} \cite{binbeshr2021systematic}. An authentic hadith is believed to belong to the prophet with a very high degree of certainty. A good hadith is also believed to belong to the prophet but with a lower degree of certainty. A weak hadith is one with no good evidence of belonging to the prophet. Finally, a fabricated hadith is believed not to belong to the prophet with a high degree of certainty.  As shown in figure \ref{fig:hadith_sample}, a hadith consists of two main parts called \textit{Isnad} (also called \textit{Sanad}) and \textit{Matn}. Isnad represents the chain of narrators who reported the hadith until it reached the scholar who compiled a collection of hadiths. The compiler judges the authenticity of a hadith and decides to include it in this collection or not based on many factors such as the trustworthiness of the narrators in the chain, the likelihood that these people have actually met each other, and how good their memory is reported to have been \cite{binbeshr2021systematic}. The second part, the \textit{Matn}, is the actual quote from the Prophet or a story about an action he took that has been passed on from narrator to narrator. A hadith can have multiple variants because it might have been transferred through multiple chains of narrators (i.e. it has multiple Isnads) \cite{azmi2019computational} and because there might be slight -- or major -- differences in how the narration itself is worded (i.e. it has multiple Matns).

The principles included in Hadith are core values for Islamic communities that are highly respected and followed by many individuals. 
\change{Many politicians and public figures are aware of this fact and have been utilizing it. For example, Barack Obama at the National Prayer Breakfast event in 2009 quoted the Hadith \textit{`None of you truly believes until he wishes for his brother what he wishes for himself’}. Joe Biden in 2020 quoted the Hadith \textit{'Whoever among you sees a wrong, let him change it with his hand; and if he is not able to do so, then let him change it with his tongue; and if he is not able to do so, then with his heart.'} promising the Muslims he would end Trump's Muslim ban after his election. In addition, most leaders of Muslim countries quote hadith in their speeches when addressing their people.
}{#change0}

On a wider scale, Hadith is also commonly used among online users with Islamic knowledge as a strong weapon for argumentation and supporting one's position on a debate. \citet{boutz2017quoting} demonstrate the utilization of Hadith in this context by examining message boards of two Arabic news websites where participants repeatedly do this practice. \change{\citet{sauda2020one} studies qualitatively an online program called \texttt{One Day One Hadith} that was launched in Indonesia. One of the organizers of this program posts a hadith every day in online communities accompanied with an article involving a discussion of the hadith as a reference to recent issues in the society. \citet{slama2017social} focuses on celebrity preachers and observes a popular way of spreading their message called \texttt{Twitter Lesson} in which they tweet a Hadith and then explain it in subsequent tweets. He presented this as one of the techniques that enhances their authority over their audience. In Bangladesh, religious content creators and public speakers learn Hadith to enrich their content \cite{rifat2022putting, rifat2022situating} and they read it in Arabic not in Bangla.}{#change1} \citet{boutz2019exploiting} investigate how ISIS used Hadith for recruitment and legitimacy assertion. They conclude that the quotation of Hadith by ISIS in their propaganda conveys to the reader a feeling of religious seriousness and commitment. They also find that ISIS was using different hadiths based on the language of the target audience to attract their attention based on their interests.

\begin{figure}
\centering
\begin{subfigure}{.33\textwidth}
  \centering
  \includegraphics[width=\linewidth]{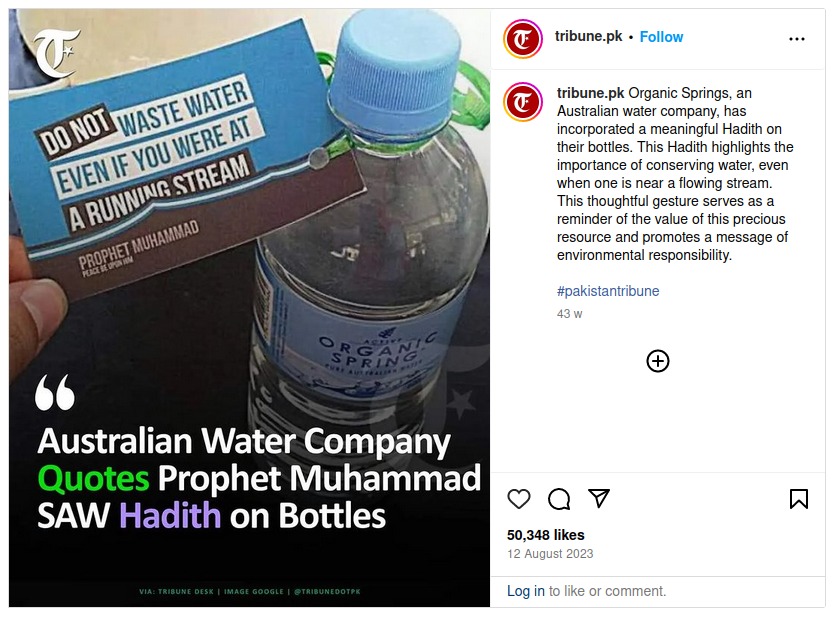}
  \caption{Hadith on bottles in Australia}
  \label{fig:social_media_1}
\end{subfigure}%
\begin{subfigure}{.33\textwidth}
  \centering
  \includegraphics[width=\linewidth]{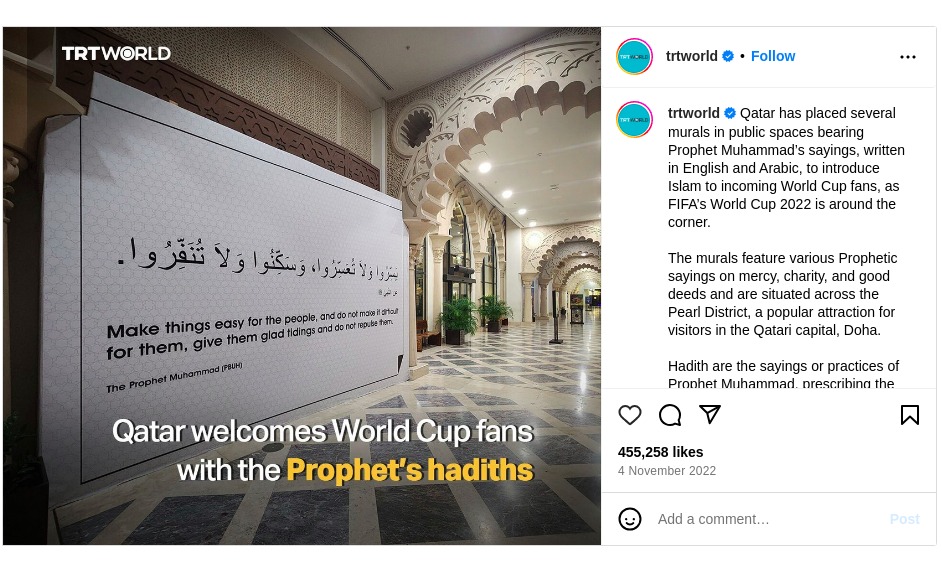}
  \caption{Hadith in the World Cup}
  \label{fig:social_media_2}
\end{subfigure}
\begin{subfigure}{.33\textwidth}
  \centering
  \includegraphics[width=\linewidth]{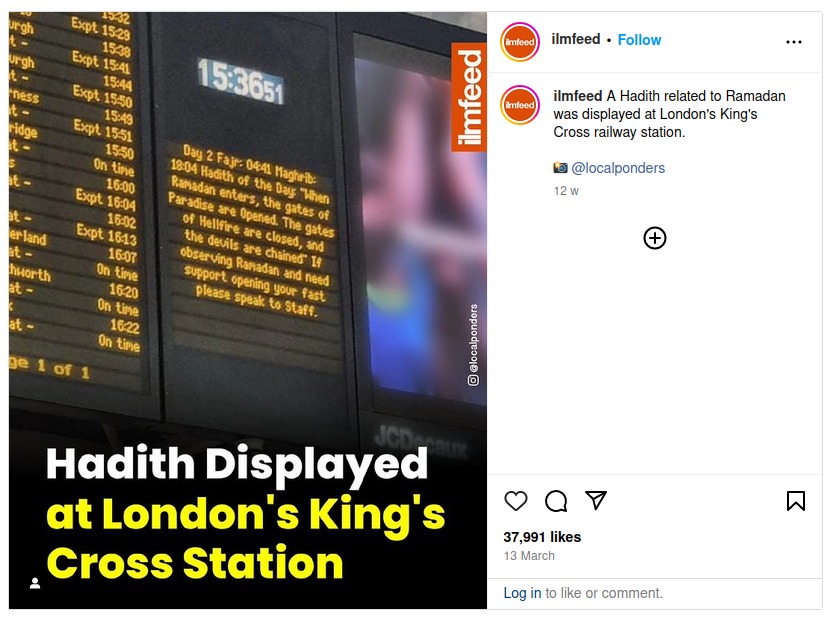}
  \caption{Hadith in London's train station}
  \label{fig:social_media_3}
\end{subfigure}
\caption{Examples of occurrences at which Hadith went viral on Instagram}
\label{fig:social_media}
\Description{Examples of occurrences at which Hadith went viral on Instagram}
\end{figure}

\change{There have been many incidents where Hadith went globally viral on social media creating discussion among users in some cases. Figure \ref{fig:social_media} shows examples of these incidents, which we describe in the following:
\begin{itemize}
    \item A photo of a water bottle with attached label containing a hadith: \textit{`Do not waste water even if you were at a running stream’} has gone viral on social media with the claim that an Australian water company sells its bottles with this label to raise awareness about consuming water efficiently (figure \ref{fig:social_media_1}). 
    The photo got admiration from many people including Muslims, but still some Muslims found it disrespectful to use religion for profit or to place religious texts in a position where they will eventually get disposed. Even though the original story of the photo was clarified in 2013 by the professional fact-checking site \texttt{Fatabyyano}\footnote{ An article clarifying the story of the hadith on bottles: \url{https://fatabyyano.net/australia/}}, many people are still circulating the photo from time to time until 2024. This raises questions about whether people are less likely to validate content if it includes appealing religious content and how religion can be misused to promote misinformation.
    \item Qatar used murals during the World Cup in 2022 with hadiths to introduce the Islamic culture. Photos of these murals went viral on social media, reaching around 500K interactions on one of those posts as shown in Figure \ref{fig:social_media_2}. A lot of people liked the initiative, but some people also considered it a strategy for Islamizing the World Cup.
   \item Recently, one of the railway networks in the UK celebrated Ramadan 2024 by showing multiple hadiths on the departure boards at King's Cross station in London.
    At the beginning, the messages mainly attracted Muslims who shared photos of the boards and expressed their gratefulness to the diversity and the inclusion in the UK. However, one of the hadiths displayed made things more intense and many non-Muslims took it to social media and complained that the message generates resentment and that preaching during stressful mornings is counter-productive.This led the company later to remove these messages. Many news websites such as \texttt{BBC}\footnote{\url{bbc.com/news/uk-england-london-68617438}}, \texttt{Daily Mail}\footnote{ \url{dailymail.co.uk/news/article-13214691/Kings-Cross-station-faces-backlash-Islamic-message-appears-board.html}}, and \texttt{The Telegraph}\footnote{\url{telegraph.co.uk/news/2024/03/20/investigation-launched-kings-cross-station-ramadan-messages}} reported the issue starting from its spark until the hadith was removed.
\end{itemize}
}{#change2}

These incidents represent a very small sample of the everyday extensive usage of Hadith in \change{communities where Muslims are present}{#change0}. They are meant to demonstrate how central this religious scripture is to the Arabic and Islamic culture. Studies that investigate Hadith through this lens are limited. We aim to fill this gap by applying a large analysis on how Hadiths are shared on social media. 

\subsection{Quantitative Studies of Hadith}

Due to the precious value of Hadith as a lingual and a cultural resource for the Arabic language, its material was digitized as early as the \finalChange{late 1970s}{#final5} \cite{al1991note} and a lot of researchers have attempted to automate the tasks associated with it. From the period of 2016-2023, there are multiple works that perform extensive literature reviews of such attempts. We pick four of them covering most of the aspects of these works to avoid redundancy \cite{saloot, azmi2019computational, binbeshr2021systematic, sulistio}.

The first thing we notice is that the most popular tasks are hadith classification according to topic and according to authenticity \cite{kabi2005, alkhatib2017rich, ghazizadeh2008fuzzy}. A clear limitation of most studies is that they do not consider the multi-label nature of Hadith in terms of topics except for few works like \cite{harrag2009neural}. Applying machine learning methods on Hadith is always motivated by the goal of having an intelligent system that can assist Islamic scholars in their studies or assist ordinary Muslims with their inquiries about Islam. We question the claim that this form of applying classification or authentication actually serves this goal in section \ref{sec:discussion}.

The second important note is that most studies perform their experiments on separate data collections that they create themselves. There is no standard dataset that allows a straightforward comparison of results. In addition, these separate data collections are relatively small with a total number of records in the order of a few hundred \change{\cite{kabi2005, harrag2009neural, ismail20132d, shatnawi2011verification, harrag2009experiments, binbadia2010itree}}{#change15}. 

Another important aspect that we also discuss in detail in section \ref{sec:discussion} is the difficulty of the task of separating the Sanad and the Matn. For example, \citet{jbara2009knowledge} performed that task manually as a preprocessing step before training their classifier. The works that attempt this task through mining approaches report results with low accuracy \cite{harrag2011, harrag2014}.

There are two more works that are not covered in the literature reviews and we find important. The first one is the LK Corpus dataset \cite{altammami} which is a dataset of 39,038 hadiths (34,088 records are available on the repository \footnote{LK Corpus: \url{github.com/ShathaTm/LK-Hadith-Corpus}}) in Arabic and English with their authenticity scores, the books they were extracted from, and the book chapter which is the field often used for topical classification. The dataset also has two separate fields for the Matn and Isnad. The second work \cite{gaanoun} creates another dataset of 26,561 hadiths (MAHADDAT) by extracting authentic hadiths from LK corpus and adding extracted fabricated hadiths from multiple Islamic websites. Afterwards, they try different classification methods including BERT-based models to classify hadiths from the two authentication levels.

\section{Methodology \& Setup}

\subsection{Data Collection}
In this work, we aim to explore the presence of hadiths in the everyday life of individuals.
For this purpose, we use data from the social media platform \change{Twitter (currently re-branded as X), since a large population of Arabs is present on it. This is in contrast to other social media, such as Reddit, that are popular in Western countries but have very few users in the Arab world. In addition, Twitter has previously been an excellent social media platform for studying the Arab and Muslim communities \cite{al2021atheists,albadi2019hateful,ibtasam2019my,abokhodair,fawzi2024pinocchio}. Given the latest changes to the platform's API which does not allow streaming large amount of data, we decided to}{} use a collection of tweets hosted on archive.org\footnote{Archive Project: The Twitter Stream Grab \url{https://archive.org/details/twitterstream}}, which is a random 1\% sample of the entire Twitter feed that was being collected periodically, resulting in a rich data source for learning what people have discussed on Twitter over a long time period. We downloaded all the available Tweets between January 2019 and January 2023. We then filter these Tweets using the \textit{lang} (language) field to only consider Tweets in Arabic. 

The next step was to extract tweets that include hadiths and for this we pick all the tweets that include the Arabic phrase \<قال رسول الله>, ``The messenger of God said''. This neutral phrase is used to report hadiths, whatever their authenticity or topic. Importantly, it implies that the author claims that they are reporting a hadith. Because Hadith is a core Arabic language resource, people might often borrow an expression from a hadith and alter its content without claiming that this new content is actually a hadith.
Filtering using the \<قال رسول الله> phrase results in 295,433 unique tweets.

\subsection{Challenges in Matching Tweets to Hadiths}
There is only one version of the Quran, the holy book of Islam, so it is a straight-forward operation to validate if a piece of text belongs to it or not by checking whether that piece is a substring of the Quran. It is not so simple for hadiths: although the number of authentic hadiths is around 30,000 \change{in some estimates}{#change17}, many of them have multiple variants. The total number of variants is in the order of hundreds of thousands \cite{jabbar, azmi2019computational}. More importantly, unlike the Quran, Muslims are allowed to narrate hadiths by meaning or paraphrase them\cite{nur}. This permission makes Muslims less sensitive to making spelling mistakes when narrating hadiths. Finally, our study doesn't only target detecting authentic hadiths but we are also interested in exploring the presence of fabricated hadiths on social media.

We rely on the LK Corpus and MAHADDAT datasets as a hadith reference for our study. To check whether a tweet includes a hadith in any of the two datasets, we tested several approaches. We first attempted to check if the tweet is a substring of a hadith in the datasets or if a hadith is a substring of a tweet in the collection. The limitation of the first approach was that users in many cases added comments before or after the hadith. The limitation of the second approach was that many users only tweeted a part of a hadith and not its whole text. Both approaches do not handle spelling mistakes.

\subsection{Matching Pipeline}
\label{sec:mapping}
To overcome the aforementioned challenges we combined the tweets with the Hadith datasets through the following multiple steps:

\begin{figure}
\centering
\includegraphics[width=0.5\linewidth]{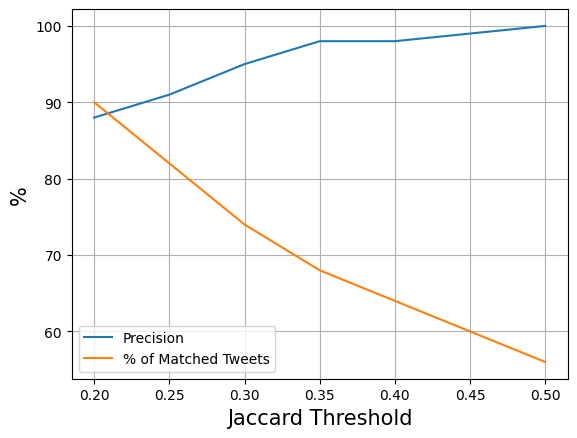}
\caption{The precision of matches and the percentage of tweets matched from the collection on different Jaccard index thresholds}
\label{fig:jaccard_threshold}
\end{figure}

\begin{enumerate}
    \item \textbf{Preprocessing:} 
    We preprocess the text of both datasets as well as our Tweets collection to remove some noise that would otherwise hinder us from matching them. Processing steps are as follows:
    \begin{itemize}
        \item \textbf{Arabic Assertion:} We remove characters that do not belong in Arabic texts (i.e. Characters outside the ASCII range of 600 and 6FF).
        \item \textbf{Text cleaning:} We remove all punctuation  since nonstandard puncation is very common on social media.
        \item \textbf{Specific Arabic Processing:} This includes removing diacritics and kashidas which are optional characters in Arabic \cite{darwish2012language}. Also, Arabic letters can be typed in multiple ways based on the word's position in a sentence. People are not always aware of the correct variant of the letter to use, so we normalize such letters to a standard form to facilitate comparison \cite{darwish2014arabic}.
        \item \textbf{Specific Hadith Processing:} A Hadith Matn usually starts with a phrase such as \textit{``The messenger of God said''}, \textit{``I heard the prophet saying''}, and \textit{``Muhammad, peace upon him, said''}. These phrases are used interchangeably by individuals to quote a hadith as they do not alter the meaning of the Matn. We remove a set of such phrases which is available in Appendix \ref{appendix_cotters}. In section \ref{sec:discussion}, we comment on the need of handling these phrase differently by computational studies on Hadith. Some fragments also exist in the middle of Matns and are often removed by individuals like \textit{``peace upon him''} and \textit{``peace upon him and his family''}, so we remove them.
    \end{itemize}
    \item \textbf{Jaccard Similarity:} We use the Jaccard index statistic as our similarity score because unlike Cosine similarity, it considers the unique set of words that appear in the texts without being sensitive to word repetition. This is relevant for our case because people often write fragments of Hadith Matns and not the entire hadiths. Considering unique words captures a simple yet compact representation of an entire hadith making it more likely to be joined to its respective fragment.
    \item \textbf{Hashing and Pairing:} Because the computation of Jaccard index is expensive on big datasets, we use the MinHash technique \cite{broder1997resemblance} implemented in the datasketch package\footnote{datasketch: \url{github.com/ekzhu/datasketch}} to compute it. We first compute the Minhash for all texts in the datasets as well as the tweet collection. Then, we compute the Jaccard coefficient between each tweet and all the hadith entries. The hadith with the highest Jaccard similarity with the tweet is softly assigned to it. The label is confirmed if the Jaccard similarity score is above a threshold value, otherwise the tweet is labeled as not matched with any of the hadiths in the datasets.
    \item \textbf{Thresholding:} To pick the threshold Jaccard score over which we consider a pair of a hadith and a tweet to match, we manually investigate random samples of 100 pairs each over a different threshold. In figure \ref{fig:jaccard_threshold}, we report the precision of matching and the coverage of the tweets collection on different thresholds. We pick the value of 0.35 as it is elbow point where diminishing returns are no longer worth the additional loss in covered tweets \cite{thorndike1953belongs}.
    \end{enumerate}

\begin{table}[]
\resizebox{\textwidth}{!}{\begin{tabular}{l l}
\hline
\textbf{Category} & \textbf{Examples of Subcategories}\\
\hline
\textbf{Knowledge (K)} &  Teaching Etiquette, Arguing \& Debate, Count of Things that only God knows\\
\textbf{Biography \& History (BH)}&  Beginning of Creation, Mention of the Prophets, Biography\\
\textbf{Jurisprudence (J)} &  Renting, Marriage, Praying, Fasting, Cases \& Ruling\\
\textbf{Interpretation (I)} &  Holy Quran, Sciences of the Holy Quran\\
\textbf{Virtues (V)} &  Virtues, Book of Qualities\\
\textbf{Asceticism (A)} &  Definition of Asceticism, Types of Asceticism, Reasons for Asceticism, Qualities of Ascetics \\
\textbf{Supplications \& Remembrances (SR)} &  Remembrances, Book of Supplications, Seeking Forgiveness, Seeking Protection \\
\textbf{Doctrine (D)} &  Islam, Faith, Temptations \& Doomsday Portents\\
\textbf{Ethics \&  Etiquette (EE)} &  Ethics, Etiquette\\
\hline
\end{tabular}}
\caption{The topical categories and examples of their subcategories on \url{sonnaonline.com}}
\Description{The topical categories and examples of their subcategories on \url{sonnaonline.com}}
\label{tab:subcategories}
\end{table}

Earlier works that address the topical distribution of Hadith usually use the chapter title of books from which the hadith is extracted. This approach is limited to analyzing a single book at a time because different compilers name similar topics differently. Moreover, it doesn't provide a clear overview of the top-level topics included in the hadith as most popular hadith books have at least 39 chapters. The most popular Hadith book \textit{Sahih Al Bukhari} \cite{bukhari1986sahih} which is used in most of earlier studies has 99 chapters. We consulted experts in Islamic studies for a source of high-level topical categorization of hadith. The most recommended source was sonnaonline\footnote{\url{http://www.sonnaonline.com}}. Sonnaonline categorizes Hadith into 9 main topics: \textit{Knowledge} \change{\textbf{(K)}}{#change22}, \textit{Biography and History} (of the prophet) \change{\textbf{(BH)}}{#change22}, \textit{Jurisprudence} \change{\textbf{(J)}}{#change22}, \textit{Interpretation} (of the Quran) \change{\textbf{(I)}}{#change22}, \textit{Virtues} \change{\textbf{(V)}}{#change22}, \textit{Asceticism} \change{\textbf{(A)}}{#change22}, \textit{Supplications and Remembrances} \change{\textbf{(SR)}}{#change22}, \textit{Doctrine} \change{\textbf{(D)}}{#change22}, and \textit{Ethics and Etiquette} \change{\textbf{(EE)}}{#change22}. We contacted the owners of the website who did not provide a structured format for the data but gave us permission to scrape their website. We scrape the available hadiths and their assigned categories \change{getting 25,997 unique hadiths with a total of 136,119 variants}{#change18} . A brief description of the content of each category is provided in section \ref{section:topical} \change{and examples of their subcategories are provided in table \ref{tab:subcategories}}{#change22}.

\change{
\subsection{Ethical and Privacy Considerations}
Religious texts are sensitive axiomatic beliefs for many people who share them and therefore the process of analyzing this practice has to be handled with care \cite{al2021atheists, wolf2024still}. Hence, an ethical approval to conduct the study was obtained from our host institution.}{#change3}

\change{
Although our main focus in this study is the nature of content shared on social media and not the users who share it, we still take precautions to maintain users' privacy. We ensured that the threads we display as examples from social media belong to news agencies or websites and not personal accounts. We also hid the comments on those threads and summarized their arguments. The data source we use doesn't provide any details about the authors of tweets except their user ID which is not anymore publicly accessible after Twitter's transformation to the X platform. Our preprocessing steps ensure removing any mentions or tags for users which might reveal their user name and hence their identity.}{#change3}

\change{
Finally, in our detailed analysis to the most interesting findings, we do not display any original threads that included Hadith but we only show the original text of the hadiths from the publicly available datasets and their English translation.
}{#change3}

\section{Results}

\begin{figure}
\centering
\includegraphics[width=0.9\linewidth]{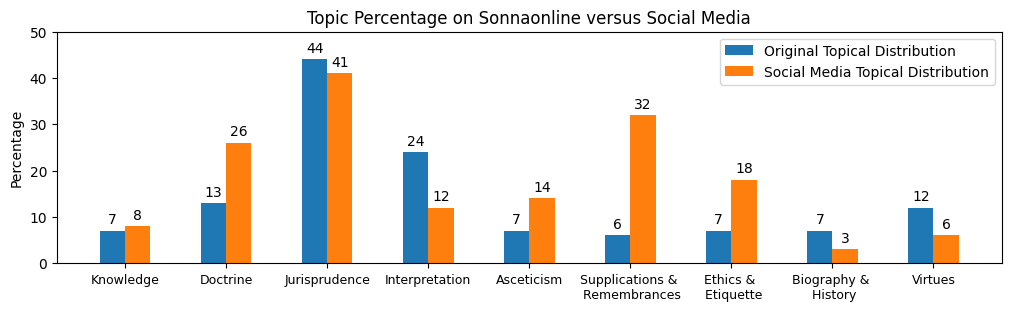}
\caption{The topical categories distribution of Hadith on Arabic social media from January 2019 to January 2023 versus on \url{sonnaonline.com} (\textbf{Note:} The sum of percentages exceeds 100 because a hadith can belong to multiple categories)}
\Description{The topical categories distribution of Hadith on \change{Twitter}{} from January 2019 to January 2023 versus on \url{sonnaonline.com} (\textbf{Note:} The sum of percentages exceeds 100 because a hadith can belong to multiple categories)}
\label{fig:topical}
\end{figure}

\subsection{Topical Distribution}
\label{section:topical}

 As shown in figure \ref{fig:topical}, there are far more hadiths on some topics than others, and some topics are over-represented or under-represented in tweets containing hadiths. The category of \textit{Jurisprudence} which addresses Islamic laws and rules is both the most frequent category of hadiths and the most frequently tweeted. The proportion of the \textit{Knowledge} category which features the value of learning and how people should use science for good deeds is roughly similar in both distributions. The rest of the categories have interesting disproportions. For example, \textit{Doctrine} which involves the philosophy behind core Islamic principals, has a portion on social media twice as big as its original portion. This can perhaps be explained by \change{the big overlap in hadiths between this category and the most over-represented category namely, \textit{Supplications and Remembrances}}{#change23}. The \change{same disproportion occurs with the}{#change23} category of \textit{Asceticism}, which calls for paying less attention to the materialistic life and more attention to spirituality. This could perhaps be explained by psychological studies that describe the human behavior of escaping stress through turning to religion \cite{livneh1996multidimensional}. \textit{Supplications and Remembrances} category \change{has a portion on social media that is 5 times bigger than its original portion. it}{#change23} includes recommended prayers and remembrances to be said in different situations like when a Muslim starts/finishes eating, when a Muslim  wakes up/goes to sleep, when it is raining, when someone is sick, or when someone dies. This category is expected to have a bigger portion than its original one since sharing content from it is thought of by Muslims as an act of religious practice. In fact, \citet{abokhodair} observed a similar pattern happening with Quran verses and they investigated it in interviews that confirmed that participants shared this content for religious practice as 'ceaseless charity' which is a concept in Islam that involves actions that keep giving a person religious credit after death. Categories with \change{relatively long hadiths}{#change23} like \textit{Interpretation}, which involves explanations of Quran verses, and \textit{Biography and History}, which includes stories about remarkable events that happened in the life of the prophet, get less attention on Social media than their original portions. \textit{Ethics and Etiquette} involves how a Muslim should behave in different situations The popularity of this category with respect to its original portion could be justified by \change{its relevance to daily situations that Muslims encounter}{#change23}. Finally, the category \textit{Virtues} involves describing how special some entities are from an Islamic lens. Entities include people, countries, particular days and months, and even foods. We don't have a hypothesis about why it is less popular than it should.

\subsection{Authenticity Distribution}

\begin{figure}
\centering
\includegraphics[width=0.8\linewidth]{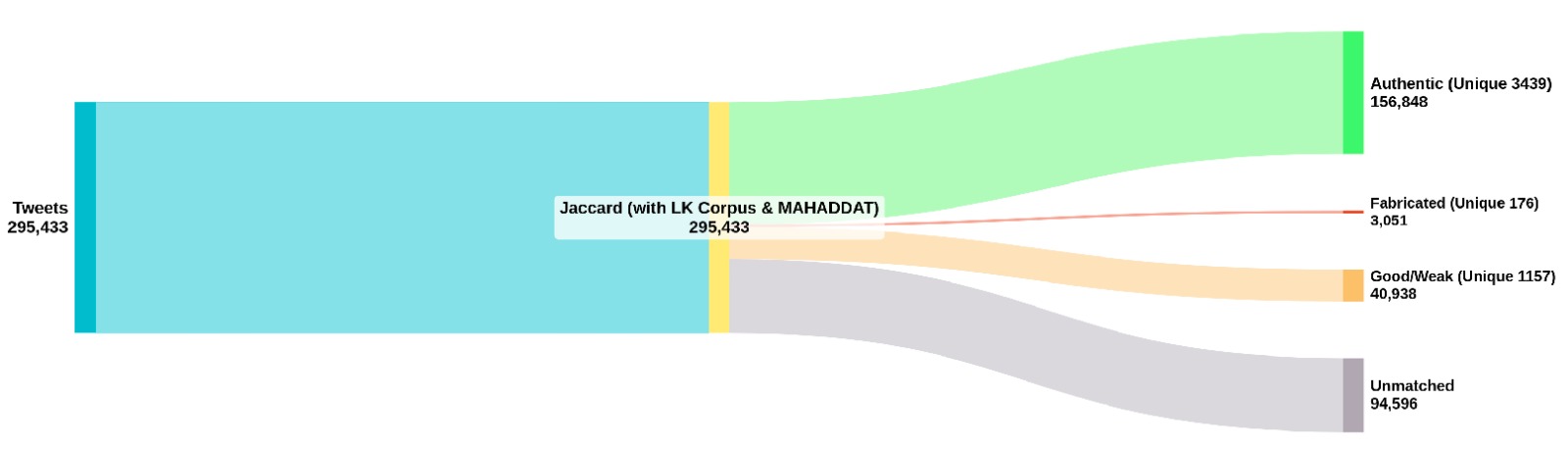}
\caption{The authenticity distribution of hadiths on Arabic social media from January 2019 to January 2023}
\label{fig:authenticity}
\end{figure}

The Sankey diagram in figure \ref{fig:authenticity} shows that most of the shared Hadith on social media is \textit{authentic} and only less than 1\% of the Tweets are identified as \textit{fabricated}. The tweets that are neither matched as \textit{authentic} nor \textit{fabricated} have not been matched with a hadith in the MAHADDAT dataset. Some of those were matched with the LK Corpus that includes, in addition to the \textit{authentic} Hadith in MAHADDAT, other records of \textit{good} and \textit{weak} Hadith. We don't investigate in detail the distribution of \textit{good} and \textit{weak} hadith as it is already a soft classification that differs from a scholar to another. Works that tried to classify both reported a difficulty with such a task and lower precision compared to other levels of authenticity \cite{ghanem2016classification}.

\begin{table}[h]
\footnotesize
\resizebox{\textwidth}{!}{\begin{tabular}{lp{11cm}c}
\hline
\textbf{Count} & \textbf{Top Authentic Hadith (translated)} & \change{Categories}{#change22}\\
\hline
4419& Two words are light on the tongue, weigh heavily in the balance, and are loved by the Most Merciful One: Glorified is Allah and praised is He, Glorified is Allah the Most Great. & \change{D, SR}{#change22}\\
3685& For me to say: (Glory is to Allah,and praise is to Allah,and there is none worthy of worship but Allah,and Allah is the Most Great) is dearer to me than all that the sun rises upon (i.e. the whole world). & \change{J, SR}{#change22}\\
3411& Our Lord, the Blessed and the Exalted, descends every night to the lowest heaven when one-third of the latter part of the night is left, and says: Who supplicates Me so that I may answer him? Who asks Me so that I may give to him? Who asks Me forgiveness so that I may forgive him? & \change{D, SR}{#change22} \\
2466& Glory be to God; Praise be to God; there is no god but God; and God is most great." A version has, "The words dearest to God are four: Glory be to God; Praise be to God; there is no god but God; and God is most great. It does not matter which you say first.& \change{SR}{#change22}\\
2353& Whoever prays for Allah's blessings upon me once, will be blessed for it by Allah ten times. & \change{A, SR}{#change22}\\
\hline
\textbf{Count} & \textbf{Top Good or Weak Hadith (translated)}\\
\hline
2676&The best invocation is that of the Day of Arafah, and the best that anyone can say is what I and the Prophets before me have said: None has the right to be worshipped but Allah Alone, Who has no partner. His is the dominion and His is the praise, and He is Able to do all things.& \change{J}{#change22}  \\
2120& There is no person who says, in the morning and evening of every day: "In the name of Allah with Whose Name nothing on earth or in heaven harms, and He is the All-Seeing, All-Knowing" three times, and is then harmed by anything. & \change{SR}{#change22}\\
1440& Whoever says 'There is no god but Allah, alone, without any partner. The Kingdom and praise belong to Him and He has power over everything one hundred times a day, it is the same for him as freeing ten slaves. One hundred good actions are written for him and one hundred wrong actions are erased from him, and it is a protection from Shaytan for that day until the night. No-one does anything more excellent than what he does except someone who does more than that.& \change{SR}{#change22}\\
1315& Whosoever begins the day feeling family security and good health; and possessing provision for his day is as though he possesed the whole world& \change{A}{#change22}\\
1046& Fast the Day of Arafah, for indeed I anticipate that Allah will forgive (the sins) of the year after it, and the year before it.& \change{J}{#change22}\\
\hline
\textbf{Count} & \textbf{Top Fabricated Hadith (translated)}\\
\hline
 1078&Whoever recites Surah Ad-Dukhan during the night, his previous sins will be forgiven.& \change{\_}{#change22}\\
 308&Gabriel came to the Prophet, may God’s prayers and peace be upon him and his family, and said to him: O Muhammad, live as long as you want, for you are dead. Love whoever you want, for you will leave him, and do whatever you want, for you will be rewarded for it. Know that the honor of a believer is his night prayers, and his dignity is his abstaining from people.& \change{\_}{#change22}
\\
 260&Whoever recites Ayat al-Kursi after every obligatory prayer, \finalChange{ Allah will grant him the heart of those who are thankful, the deeds of the siddiqun, the rewards of the Prophets and will extend his right hand with Mercy and will not prevent him from entering Paradise until he dies whereupon he will be made to enter it.}{#final1}\footnote{\label{evaluationRemark}This hadith is considered \textbf{weak} and not \textbf{fabricated} by some scholars.}& \change{\_}{#change22}\\
 147&No one honors women except a generous person, and no one insults them except a mean person.& \change{\_}{#change22}
\\
 131&Pray for Allah's blessings upon me often, for God has appointed an angel for me at my grave. When a man from my people prays for Allah's blessings upon me, that angel says to me, O Muhammad, that so-and-so, the son of so-and-so, has prayed for Allah's blessings upon you at this hour.\footref{evaluationRemark} & \change{\_}{#change22}\\
 \hline
\end{tabular}}
\caption{The English translation of the most shared 5 hadiths on Arabic social media between January 2019 and January 2023 for each authenticity level}
\Description{The English translation of the most shared 5 hadiths on Arabic social media between January 2019 and January 2023 for each authenticity level}
\label{tab:most_popular_english}
\end{table}

Table \ref{tab:most_popular_english} shows the English version of the hadiths we found to be most popular in our collection\footnote{English translations from \url{sunnah.com}}. The original Arabic version of these hadiths can be found in Appendix \ref{appendix_arabic_hadith}. The top 5 hadiths on social media include 4 from the \textit{authentic} category and 1 from the \textit{Weak/Good} categories. In general most of the top shared hadiths belong to the \textit{Supplications and Remembrances} topical category. In addition to the \textit{Supplications and Remembrances}, the top and the 5th hadiths in the \textit{Weak/Good} belong to \change{\textit{Jurisprudence}}{#change23} category where they highlight the value of prayers on the day of 'Arafah' which is the day that precedes the Islamic feast "Greater Bairam". The 4th hadith under the same authenticity level belongs to \textit{Asceticism} topical category where it tells that people should feel thankful for the the for granted yet valuable things in their life like safety and health. The fabricated hadith is not labeled by our source of topical categorization; however the semantics of its top records are likely to belong to \textit{Supplications and Remembrances} except the 2nd which tends to \textit{Asceticism} and the 4th which tends to \textit{Ethics and Etiquette}.

\subsection{Time Distribution and Seasonality}
In tests with the live Twitter API, we \change{found}{#change0} that new tweets containing the query phrase \<قال رسول الله> appear at an average rate of 10 tweets per minute. This shows how hadith is a fundamental cultural resource for many Muslims. Even though the data collection we examine in this study is just a random 1\% of the Twitter feed in all languages, we still find tweets that include hadiths on each of the days in the time period covered.

\begin{figure}
\centering
\begin{subfigure}{.3\textwidth}
  \centering
  \includegraphics[width=\linewidth]{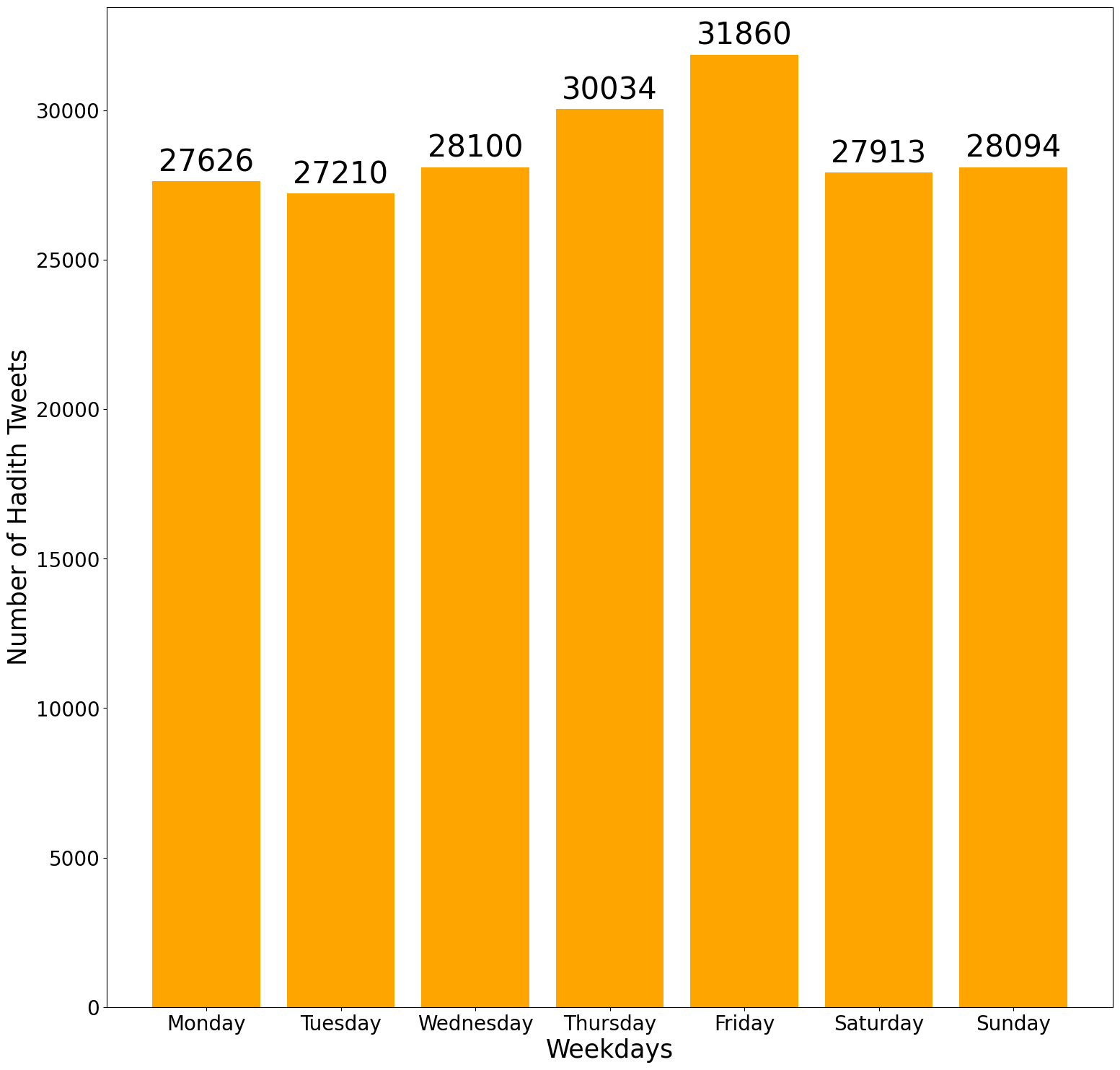}
  \caption{Weekdays}
  \label{fig:sub1}
\end{subfigure}%
\begin{subfigure}{.6\textwidth}
  \centering
  \includegraphics[width=\linewidth]{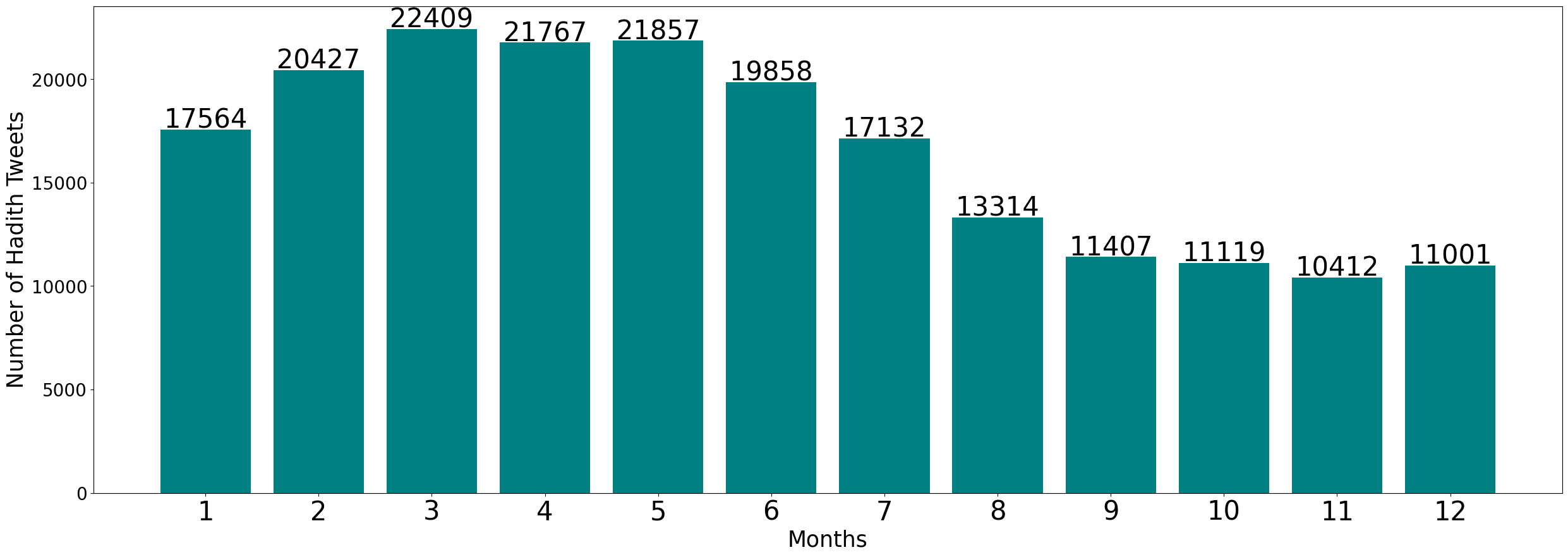}
  \caption{Months}
  \label{fig:sub2}
\end{subfigure}
\caption{The distribution of hadiths over weekdays and months. \textbf{Note:} Data for January 2023 is filtered for (b) so that all months have an equal number of occurrences.}
\label{fig:weeks_months}
\Description{The distribution of hadiths over weekdays and months. \textbf{Note:} Data for January 2023 is filtered for (b) so that all months have an equal number of occurrences.}
\end{figure}

\begin{table}[h]
\footnotesize
\resizebox{\textwidth}{!}{\begin{tabular}{lp{10cm}c}
\hline
\textbf{Gini Index} & \textbf{Most unequally distributed hadiths (translated)} & \change{Categories}{#change22}\\
\hline
0.85 & The best invocation is that of the Day of \textbf{Arafah}, and the best that anyone can say is what I and the Prophets before me have said:None has the right to be worshipped but Allah. Alone, Who has no partner. His is the dominion and His is the praise, and He is Able to do all things.& \change{J}{#change22} \\
0.8 & Fasting on the Day of \textbf{Arafah}, I hope from Allah, expiates for the sins of the year before and the year after.& \change{J}{#change22} \\
0.71 & O Allah, bless my nation in their early mornings& \change{J, EE}{#change22} \\
0.7 & The sun and the moon are two signs of Allah; they are not eclipsed on account of anyones death or on account of anyone's birth. So when you see them, glorify and supplicate Allah.& \change{J, SR}{#change22}  \\
0.69 & Fasting the day of \textbf{Ashura}, I hope, will expiate for the sins of the previous year. & \change{J}{#change22}  \\
0.66 & Seek Lailat-ul-Qadr \textbf{(Night of Decree)} in the odd nights out of the last ten nights of Ramadan. & \change{J}{#change22} \\

\hline
\textbf{Gini Index} & \textbf{Most evenly distributed hadiths (translated)} & \change{Categories}{#change22}\\
\hline
0.08 & The Messenger of Allah said "When the practice of honoring a trust is lost, expect the Last Day." Someone asked: "How could it be lost?" The prophet replied, "When the government is entrusted to the undeserving people, then wait for the Last Day."& \change{D, J, EE}{#change22}  \\
0.09 & It is from the excellence of (a believer's) Islam that he should shun that which is of no concern to him& \change{D, A}{#change22} \\
\hline
\end{tabular}}
\caption{The hadiths whose mentions on Arabic social media are the most vs. least evenly distributed in terms of days (a high Gini index indicates a hadith whose mentions spike on a specific day)}
\Description{Most and least evenly distributed hadiths on Arabic social media}
\label{tab:seasonality}
\end{table}

Figure \ref{fig:weeks_months} shows that Thursday and Friday have more tweets than the rest of the week. This is expected because Friday is a special day for Muslims that has an entire chapter in the Quran \cite{abdel2020general} and a special prayer called "Friday prayer" \cite{goitein2010v}. This also explains the higher number of tweets on Thursdays: In Islam, the religious day starts at sunset on the previous calendar day, so from a religious perspective, Friday begins when the sun sets on Thursday. The figure also shows a higher number of tweets between March and May compared to the rest of the year. This is also expected because Ramadan, the special month of Muslims that they are required to fast \cite{alkandari2012implications} and become more religious in, occurred from 2019 to 2022 during these months. Ramadan time is determined based on the Islamic lunar calendar which has on average 11 less days than the Gregorian calendar. As a result, Ramadan moves backward each year \change{\cite{al2019study}}{#change19} and we expect the maximum volume of hadith sharing to move similarly in the coming years.

We inspect unusual spikes and seasonality in the number of times that hadiths are mentioned by social media users. We use the Gini coefficient \cite{gini1912variability} to measure, for each hadith, the statistical dispersion (or inequality) in the distribution of the number of tweets mentioning that hadith per day. A coefficient close to 0 means that this hadith was shared in similar numbers on each day in the time period covered; a coefficient of 1 would mean that a hadith is only shared on a single day. We limit the analysis to the 455 hadiths that were tweeted 100 times or more. As shown in table \ref{tab:seasonality}, the hadiths with high Gini coefficients belong to special Islamic \change{days}{#change0} like 'Arafah' day, Ashura day, and the Night of Decree which as a special night within the special month of Ramadan. Interestingly, the 4th most seasonal hadith involves the Islamic instructions on what to do during lunar and solar eclipse which are indeed seasonal events.
\change{We investigate in a similar way the Gini indices of hadiths over weekdays (Table \ref{tab:seasonality_week}) and over months (Table \ref{tab:seasonality_month}). Friday doesn't only include the highest Hadith traffic over the week but we also found that the most seasonal hadiths on weekly basis are related to it and the recommended forms of religious practice within it. When it comes to months, seasonal hadiths are about multiple months in the Islamic lunar calendar. For example, the first hadith mainly demonstrates the value of fasting during the month of Muharram, which also includes the seasonal Ashura day. The second hadith is about the value of praying at night during Ramadan, the third hadith is about the value of righteous actions during the first 10 days of the month of Dhul-Hijjah, and the fourth hadith is about the value of fasting 6 days during the month of Shawwal. }{#change28}

\begin{table}[h]
\footnotesize
\resizebox{\textwidth}{!}{\begin{tabular}{lp{10cm}c}
\hline
\change{\textbf{Gini Index}}{#change28}& \change{\textbf{Most Seasonal Hadith in terms of Weekdays (translated)}}{#change28} & \change{Categories}{#change28}\\
\hline
\change{0.78}{#change28} & \change{There is a time on \textbf{Friday} at which no Muslim would ask Allah for what is good but He would give it to him.}{#change28}& \change{J, SR}{#change22} \\
\change{0.75}{#change28} & \change{It was \textbf{Friday}  from which Allah diverted those who were before us. For the Jews (the day set aside for prayer) was Saturday, and for the Christians it was Sunday. And Allah turned towards us and guided us to \textbf{Friday} for us. In fact, He made \textbf{Friday}, Saturday and Sunday. In this order would they (Jews and Christians) come after us on the Day of Resurrection. We are the last of among the people in this world and the first among the created to be judged on the Day of Resurrection.}{#change28}& \change{J}{#change22} \\
\change{0.72}{#change28} & \change{When it is \textbf{Friday} , the angels stand at every door of the mosque and record the people in the order of their arrival, and when the Imam sits (on the pulpit for delivering the sermon) they fold up their sheets and listen to the mention (of Allah).}{#change28} & \change{\_}{#change22} \\

\hline
\change{\textbf{Gini Index}}{#change28}& \change{\textbf{Least Seasonal Hadith in terms of Weekdays (translated)}}{#change28} & \change{Categories}{#change28}\\
\hline
\change{0.01}{#change28} & \change{Our Lord, the Blessed and the Exalted, descends every night to the lowest heaven when one-third of the latter part of the night is left, and says: Who supplicates Me so that I may answer him? Who asks Me so that I may give to him? Who asks Me forgiveness so that I may forgive him?}{#change28}& \change{D, SR}{#change22}  \\
\change{0.015}{#change28}& \change{Two words are light on the tongue, weigh heavily in the balance, and are loved by the Most Merciful One: Glorified is Allah and praised is He, Glorified is Allah the Most Great.}{#change28}& \change{D, SR}{#change22}  \\
\hline
\end{tabular}}
\caption{Most and least seasonal hadiths in terms of weekdays (a high Gini index indicates a hadith whose mentions tend to spike on the same day of the week). Special Islamic days/months are in \textbf{bold} - emphasis ours.}
\Description{Most and least seasonal hadiths on Arabic social media in terms of Weekdays}
\label{tab:seasonality_week}
\end{table}

\begin{table}[h]
\footnotesize
\resizebox{\textwidth}{!}{\begin{tabular}{lp{10cm}c}
\hline
\change{\textbf{Gini Index}}{#change28}& \change{\textbf{Most Seasonal Hadith in terms of Months (translated)}}{#change28} & \change{Categories}{#change28}\\
\hline
\change{0.87}{#change28} & \change{The best month for fasting next after Ramadan is the month of Allah, the \textbf{Muharram}; and the best prayer next after the prescribed prayer is prayer at night.}{#change28}& \change{J}{#change22} \\
\change{0.81}{#change28} & \change{Whosoever performs prayers at night during the month of \textbf{Ramadan}, with Faith and in the hope of receiving Allah's reward, will have his past sins forgiven.}{#change28}& \change{J, V}{#change22} \\
\change{0.80}{#change28} & \change{The Messenger of Allah said, "There are no days during which the righteous action is so pleasing to Allah than these ten days (i.e., the first ten days of \textbf{Dhul-Hijjah})." He was asked: "O Messenger of Allah, not even Jihad in the Cause of Allah?" He replied, "Not even Jihad in the Cause of Allah, except in case one goes forth with his life and his property and does not return with either of it."}{#change28} & \change{J}{#change22} \\
\change{0.79}{#change28} & \change{He who fasts in the month of Ramadan, and also six days in the month of \textbf{Shawwal}, it is as if he fasted for the whole year.}{#change28} & \change{J}{#change22} \\

\hline
\change{\textbf{Gini Index}}{#change28}& \change{\textbf{Least Seasonal Hadith in terms of Months (translated)}}{#change28} & \change{Categories}{#change28}\\
\hline
\change{0.04}{#change28}& \change{Two words are light on the tongue, weigh heavily in the balance, and are loved by the Most Merciful One: Glorified is Allah and praised is He, Glorified is Allah the Most Great.}{#change28}& \change{D, SR}{#change22}  \\
\change{0.043}{#change28} & \change{For me to say: (Glory is to Allah,and praise is to Allah,and there is none worthy of worship but Allah,and Allah is the Most Great) is dearer to me than all that the sun rises upon (i.e. the whole world).}{#change28}& \change{J, SR}{#change22}  \\
\hline
\end{tabular}}
\caption{Most and least seasonal hadiths in terms of months  (a high Gini index indicates a hadith whose mentions tend to spike in the same month of the year). Special Islamic days/months are in \textbf{bold} - emphasis ours.}
\Description{Most and least seasonal hadiths on Arabic social media in terms of months  (a high Gini index indicates a hadith whose mentions tend to spike in the same month of the year). Special Islamic days/months are in \textbf{bold}.}
\label{tab:seasonality_month}
\end{table}

The least seasonal hadiths on the other hand involve general topics like how dangerous it is when unqualified people are in charge of something, how a Muslim should not try to know information that is not of his business, and that praying at night in general is loved by God. \change{The least seasonal hadiths on weekly and monthly basis are actually the same hadiths found in table \ref{tab:most_popular_english} which appear most often on social media in general.}{#change28}

\section{Discussion}
\label{sec:discussion}

This study is an attempt to explore the spread of Hadith as an influential Arabic language resource among individuals on Arabic social media and to address religious expression as a form of self expression. In this section, we discuss the study's findings, its implications and limitations.


\subsection{General Findings}

\textbf{Hadith Presence on Social Media:} Regarding \textbf{RQ2} and \textbf{RQ3}, Hadith is generally quite common on Arabic social media. The most overrepresented topical categories of Hadith on social media are \change{\textit{Supplications and Remembrances}, \textit{Ethics and Etiquette}, \textit{Asceticism}, and \textit{Doctrine}}{#change23}, whereas the most underrepresented topical categories are \textit{Interpretation} and \change{\textit{Biography and History}}{#change23}. In most cases, quoted Hadith is \textit{authentic}; however, many \textit{fabricated}, \textit{good}, and \textit{weak} Hadith also get quite popular. Hadith presence is at its most around Friday in terms of weekdays and around Ramadan in terms of months due to their special value for Muslims. Hadiths that have seasonal presence are usually linked with seasonal events like solar/lunar eclipse, Arafah day, and Ashura day. Hadiths that have steady presence discuss general topics that are relevant at any time.

\begin{figure}
\centering
\includegraphics[width=0.6\linewidth]{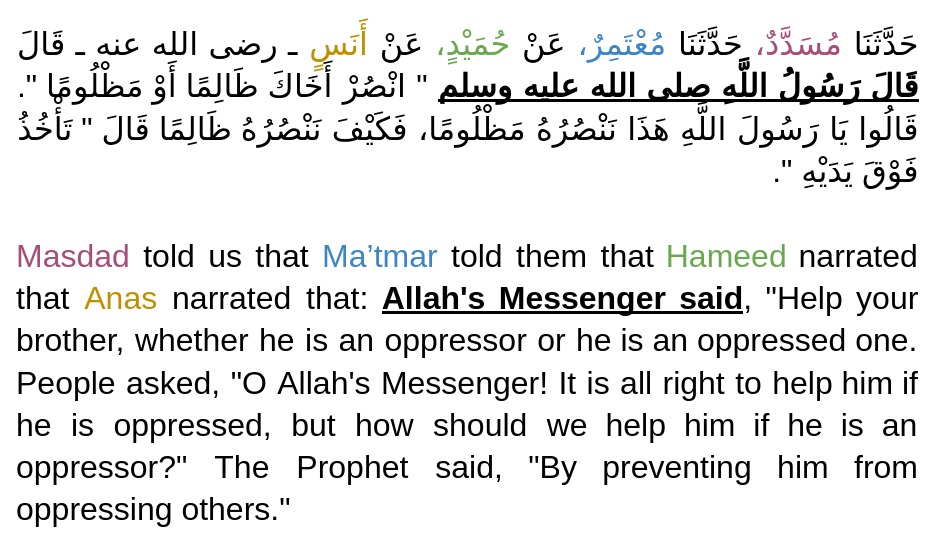}
\caption{The same Hadith presented in figure \ref{fig:hadith_sample} with the keyring phrase written in bold and underlined.}
\label{fig:cotter}
\end{figure}

\textbf{Gaps in quantitative studies:} In response to \textbf{RQ1}, even though LK Corpus \cite{altammami} and MAHADDAT \cite{gaanoun} are valuable works compared to earlier studies that analyze Hadith  quantitatively, we believe there are limitations that need to be addressed in future works that attempt to construct datasets for Hadith or classify Hadith in terms of authenticity. LK Corpus has 2 fields for authenticity (English\_Grade and Arabic\_Grade). The first one has 42 unique values while the second one has 296. Values are not only inconsistent, but they also cannot be mapped to the 4 standard known authenticity levels because many of the records include all the 4. For MAHADDAT, we believe that the classification task should consider the practical contexts where \textit{authentic} and \textit{fabricated} hadiths are quoted. We reproduced the classification experiment with the highest accuracy using BERT, however, on repeating the experiment with preprocessed texts of Hadith using our pipeline, the accuracy is dropped by 5\%. Our hypothesis was that the way Hadith is written in books is totally different than the way it is written on websites where individuals ask Islamic scholars whether a hadith is \textit{authentic} or not (which is the source of the \textit{fabricated} class in the study). For validation, we investigated the percentage of the presence of the phrase \textit{The messenger of God said} in records of both classes and we find that it is present in 16.3\% of the \textit{authentic} class records compared to 2.6\% of the \textit{fabricated} class records.
We propose formalizing a set that includes the phrases we removed in section \ref{sec:mapping} like the one shown in figure \ref{fig:cotter} and to handle them based on the application while processing Hadith computationally. These parts of Matn act as keyrings in a key chain. They don't belong to the chain of narrators (Isnad) but they don't belong to the key message of a hadith either. They exist to allow smooth transitioning from the Isnad to the Matn. We include a primary insufficient set of such phrases in Appendix \ref{appendix_cotters}.

There is an annual increase in papers about Hadith \cite{azmi2019computational}; However, literature reviews still acknowledge the need for robust standard datasets of Hadith \cite{azmi2019computational, binbeshr2021systematic}. Despite the presence of many interesting ideas within earlier studies like investigating relations of Hadith vector spaces \cite{ismail20132d}, relations between keywords in Hadith chapters \cite{nohuddin2015keyword}, prophetic medicine \cite{alrumkhani2016tibbonto}, and networks of Hadith narrators that resemble social media networks \cite{csenturk2005narrative}, the big bulk of quantitative Hadith research is directed towards Hadith classification and authentication without a worthy objective. Classifying or authenticating Hadith records in literature books is not so practical because gold labels are available for all possible instances already. Models trained for such tasks are not helpful for Islamic scholars or Muslim individuals as claimed because they don't cover the challenging context where Hadith could be present in fragmented forms like social media, official talks, or press. A system with a friendly GUI like the one introduced by \cite{hamam2015data} is sufficient for handling books.

\textbf{Gaps in qualitative studies:} The existence of studies that examine the intersection of Hadith with emerging social phenomena under the umbrella of digital religion is a decent step towards understanding human behavior. Nevertheless, we think that upgrading these studies through using big data will make them much more effective. For instance, the attempt of \citet{lucas2008major} to see the major topics of Hadith through a western eye could be guided by big data to pick a sample of Hadith that is more popular among Muslims. None of the hadiths picked by him belonged to the top hadiths in our findings. Similarly, the work of \citet{boutz2019exploiting} to examine how ISIS used Hadith for recruitment and legitimacy could have another dimension of including how Hadith and ISIS intersect among individuals. In our tweets collection from 2019 to 2023 after ISIS has already been defeated, all of the tweets that included Hadith in addition to the word 'ISIS' was criticizing its ideology and practices. If our pipeline is applied to the data in 2015 when ISIS was most popular, a broader view on how recruitment to ISIS was being done among individuals can be achieved. 

\subsection{Implications}

Our findings \change{address one of the clear gaps identified by \citet{wolf2024still} that there is a lack in the HCI and CSCW research on religion and spirituality that contributes knowledge necessary for understanding the human social behavior. This is necessary because global events like COVID-19 caused a lot of religious expression activities to  be migrated to online communities leading to a new set of socio-religious interactions and urging the research community to investigate them \cite{aduragba2023religion, claisse2023keeping}.}{} Our work fills part of this gap by studying Hadith, a key Arabic language resource, through a social lens. Our work contributes to the previous works in this direction, \change{such as the CSCW work of \citet{abokhodair} on Quran sharing on Twitter, and draws the attention of}{} social media platform designers as well as Islamic applications/websites creators \change{towards aspects of online communication that have previously been given little attention}{}.

\change{Our analysis showed the importance of hadiths for social media users, with interest in specific hadiths spiking at different times such as during Ramadan and on Fridays. Application designers are encouraged to take this into account and develop new features that allow users special means of religious expression during these times. They can trace times of the day when people tend to share or read Hadith and use notifications and other creative ideas to increase the engagement with their products. They can also include the Hijri (Islamic lunar) calendar along with the Gregorian one where it is relevant and mark the special religious days for Muslim users.}{} 

Social media platforms should pay attention to religious misinformation and train algorithms to warn users about it. Though it may only be a limited portion of all hadiths shared, it can be influential in manipulating religious users \change{\cite{boutz2017quoting, boutz2019exploiting}}{#change4}. \change{
Both platform designers and researchers should pay attention to how fabricated hadiths spread on social media, who spreads and refutes them, and how the characteristics of Arab users interacting with fabricated hadiths align with (or differ from) the Arab users interacting with fake news \cite{fawzi2024pinocchio}. Such research in misinformation shall be of high importance, especially when linked to how extreme groups might use religious misinformation to promote their agendas \cite{boutz2019exploiting}.}{}
Islamic application designers should consider developing intelligent systems that can validate the authenticity of hadiths within a social media thread or an online article and not just those coming from literature books.

Finally, for researchers who try to understand the role of Hadith in driving different social phenomena \change{\cite{slama2017social, boutz2017quoting}}{#change0}, our results show that there is great potential provided by big data to draw much more general and supported conclusions \change{than what would be possible in small-scale qualitative work}{#change0}.

\subsection{Limitations \& Future Work}
\label{sec:limitations}


\finalChange{A limitation of this study is that the analysis relies primarily on Twitter Archive's data, which consists of a random stream of tweets not specifically curated for examining the Hadith topic. Although this approach yielded a large number of tweets containing Hadith, a more targeted collection through a filtered stream focusing explicitly on Hadith-related content could have potentially captured a greater volume of relevant tweets. However, recent constraints imposed by the Twitter API limited our ability to access such filtered streaming data, making this data dataset our most viable option for this study.}{walid}
In addition, we focused our study on hadiths shared in Arabic. \citet{sulistio} show that there is a growing interest in studying hadiths in Malay, Indonesian, English, and Urdu. A similar study on such languages would contribute to a better understanding of the role of hadiths in different societies. Another limitation was that we analyzed only around 70\% of the extracted Tweets to maintain matching accuracy. The presence of more efficient mining techniques for Hadith Matn within social media threads or press articles would have enabled the analysis of a larger portion of the collection. An important aspect of our tweets collection is that it only considers hadiths in text format. People, however, also share hadiths in images, videos, and external links.
Finally, we only study data from a single social media platform, namely X (formerly Twitter).
Other social media platforms offer other forms of communication where Hadith is popular like stories and reels on Whatsapp, Facebook, and Instagram as well as community chats that are getting popular on Telegram as well as Whatsapp \change{\cite{sauda2020one}}{#change0}.

\change{A natural extension to our work would be to conduct a qualitative study similar to \cite{ibtasam2019my, abokhodair, ibrahim2024tracking}. Responses to surveys as well as interviews would provide a clearer view on the motives for sharing Hadith in general and specific hadiths in particular. It shall also validate the quantitative findings of our study through the contexts of the lives of people who share Hadith.}{#change27}


\section{Conclusion}
In this work, we characterized and explored many aspects of Hadith presence on Arabic social media. We managed to craft a dataset of around 300,000 tweets that include Hadith and to successfully overcome the challenges of joining it with other datasets that allow gaining insights on Hadith's authenticity distribution, topical category distribution, and seasonality. We find most Hadith on social media to be authentic and we observe a disproportionality between the actual topical categories distribution of Hadith and its respective presence on social media. We also identify seasons when Hadith is most popular and the most seasonal hadiths on social media. We highlight the shortages present in recent quantitative studies that address Hadith and show the potential of upgrading qualitative studies that involve Hadith as well. We hope that these findings will motivate further works  to explore digital religious expression and to offer creative ideas for designing platforms that handle religious misinformation and boost spirtual expression.

\bibliographystyle{ACM-Reference-Format}
\bibliography{references}

\appendix

\section{Keyrings of Hadith}
\label{appendix_cotters}


\centering
\includegraphics[width=0.8\linewidth]{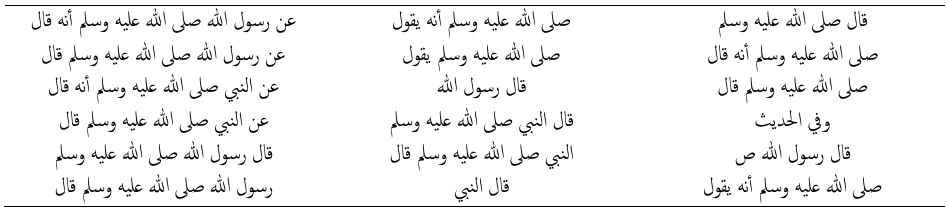}

\section{Most popular Hadith in Arabic}
\label{appendix_arabic_hadith}

\centering
\includegraphics[width=\linewidth]{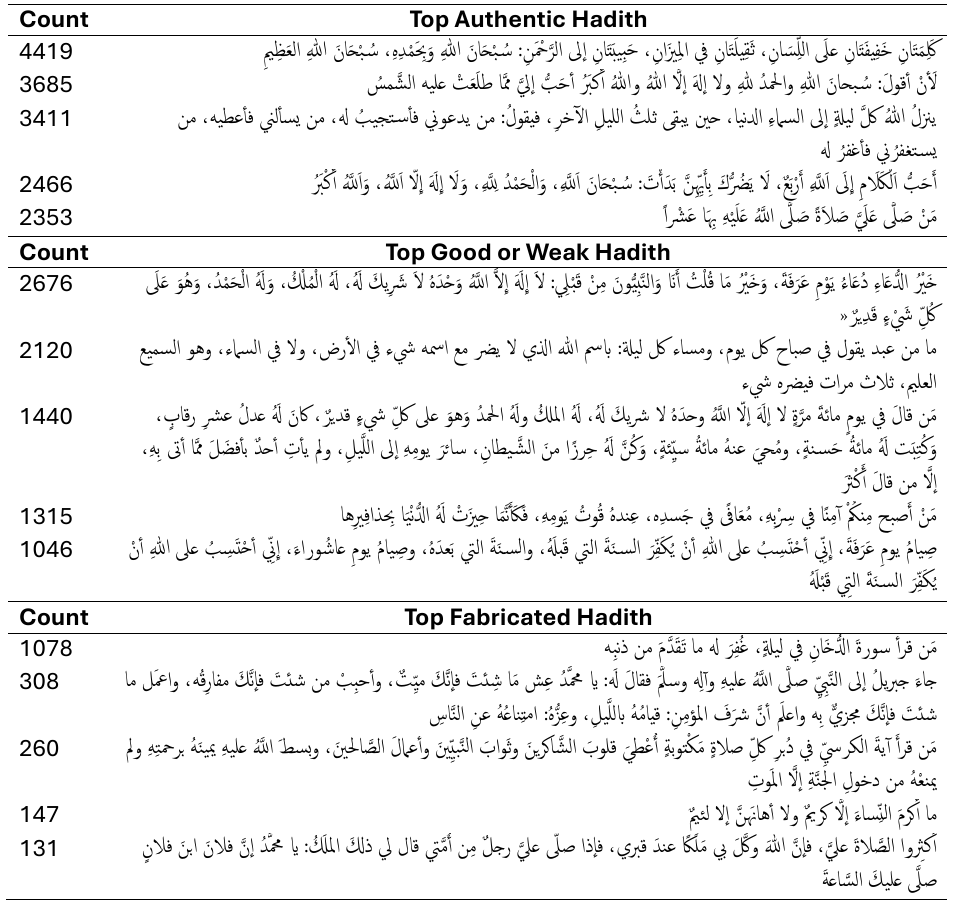}

\end{document}